\documentclass[12pt]{article}
\usepackage{graphics, color}
\usepackage{amssymb}
\usepackage{amsmath}

\begin{document}

\baselineskip=15.5pt
\pagestyle{plain}
\setcounter{page}{1}


\begin{titlepage}
\bigskip
\centerline{\Large \bf Wrapping Interactions and the Konishi Operator}
\bigskip\bigskip\bigskip

\centerline{\large Cynthia A. Keeler$^a$ and Nelia Mann$^b$}
\bigskip
\centerline{\em $^a$ Department of Physics, University of California} \centerline{\em Berkeley, CA 94702} \centerline{\em ckeeler@berkeley.edu}
\smallskip
\centerline{\em $^b$ Enrico Fermi Institute, University of Chicago} \centerline{\em Chicago, IL  60637} \centerline{\em nelia@theory.uchicago.edu}
\bigskip
\bigskip
\bigskip\bigskip


\begin{abstract}
\medskip
We present a calculation of the four-loop anomalous dimension of the $SU(2)$ sector Konishi operator in $\mathcal{N} = 4$ SYM, as an example of ``wrapping'' corrections to the known result for long operators.  We use the known dilatation operator at four loops acting on long operator, and just calculate those diagrams which are affected by the change from operator length $L > 4$ to $L = 4$.  We find that the answer involves a $\zeta[5]$, so it has trancendentality degree five.  Our result differs from previous proposals and calculations.  We also discuss some ideas for extending this analysis to determine finite size corrections for operators of arbitrary length in the $SU(2)$ sector.
\noindent
\end{abstract}
\end{titlepage}

\section{Introduction}

One of the most important recent advances in $AdS/CFT$ has been the development by Beisert, Eden, and Staudacher of the Bethe Ansatz solution to the problem of finding the anomalous dimensions of long single-trace operators in $\mathcal{N} = 4$ SYM at large $N$ \cite{BES}.  The solution depends on identifying a long operator as a spin chain, where the sites on the chain are identified with fields in the operator, an approach first pioneered in \cite{MZ}.  (A selection of important papers produced along the way is \cite{background}.)  Along this spin chain move magnons, impurities in the operator, which each carry momentum and charge.  The anomalous dimension of the operator is then identified with the energy of the spin chain; it is just the sum of the energies of individual magnons.  Finally, the momenta of the magnons are quantized by a Bethe equation which depends on the S-matrix for interactions between magnons.

These equations have proven very successful.  They naturally split the S-matrix into two parts, a tensor structure which is determined by the global symmetry of the spin chain system, and a dressing phase, which is constrained by crossing relations.  The dressing phase first affects the anomalous dimensions at four loops and contains in it any trancendentality that the anomalous dimensions have; the tensor structure part of the S-matrix alone always leads to algebraic results.  This S-matrix has been successfully tested up to four loops, first by a four-loop calculation of the cusp anomalous dimension \cite{Dixon} and then by a four-loop calculation of the dressing phase \cite{BMR}.

However, these equations can break down when acting on operators of finite length.  Heuristically, we expect this to be the case because the S-matrix is defined between asymptotically free states, and a system on finite size allows for none.  In this system, as we expand the equations in the 't Hooft parameter of the gauge theory, the interaction range grows: at one loop interactions are nearest-neighbor, at two loops next-nearest-neighbor, and so on.  The system no longer admits asymptotically free states when the loop expansion reaches the length of the operator.  In fact, we can see the origin of this breakdown explicitly in the gauge theory.  When the loop expansion reaches the length of the operator, certain Feynman diagrams contributing to the anomalous dimension disappear, and others contribute for the first time; these are the ``wrapping'' interaction diagrams.  Some previous work on wrapping corrections appears in \cite{wrapping}.

We wish to examine this problem in the $SU(2)$ sector, partly because of its relative simplicity as compared to the full $PSU(2,2|4)$ system but also because the structure for long operators has been that once one has the solution for the $SU(2)$ sector, the equations for the other sectors can be built up from it based on the global symmetry in much the same way a Lie group is built up from $SU(2)$ roots.
Although the long-term goal must be to examine these effects for operators of arbitrary length, in what follows we will study this effect for the smallest non-trivial operator in the $SU(2)$ sector, the length $4$ Konishi operator.  The dilatation operator in the $SU(2)$ sector is already known to four loops, we will use this knowledge in our calculation to limit the number of diagrams needed.  Thus we will only calculate those diagrams present for long operators which disappear and those wrapping diagrams which are fundamentally new for $L = 4$.  This basic technique was presented in various venues in the  Fall of 2007 \cite{talks}.

The final result will be the four-loop correction to the dimension of the Konishi operator.  Various proposals have been made for this result in \cite{guesses} and \cite{BES}.  These guesses all share the feature that the trancendentality of the dimension was predicted to be no larger than that for a long operator at four loops: degree three.  In addition, while this work was in preparation a calculation of the dimension was published \cite{Milano}; this calculation gave a degree of trancendentality of five.  Our result agrees with none of the above results, although it also produces a dimension with degree five trancendentality.

The paper will be organized as follows.  In the following section we will present background material on the dilatation operator at four loops acting on long operators.  In section three we will provide a systematic description of all diagrams that contribute to the calculation.  In section four we will show the effect of those ``unwrapped, maximal length'' diagrams that are present for the long operator calculation but which disappear for the Konishi operator, and in section five we will present the effect of the wrapping diagrams, as well as the final result for the anomalous dimension of the Konishi operator.  In section six we will discuss future directions, in particular ideas we have for generalizing the analysis to arbitrary length.  Details of diagrammatic calculation, as well as a complete list of diagrams calculated, can be found in the appendix.

\section{The Long Range Hamiltonian}

The key initial insight of using integrability to diagonalize the dilatation operator in $\mathcal{N} = 4$ SYM was to express the dilatation operator as a spin chain Hamiltonian, with fields in the operator re-interpreted as sites in the spin chain.  In the $SU(2)$ sector, operators are of the type
$$
\mathcal{O}^{I} = \mbox{Tr} (X^{a_1}X^{a_2}\cdots X^{a_L})
$$
where the $a_{i} = 1,2$.  The dilatation operator is written as
$$
\mathcal{H} = \sum_{\ell = 0}^{\infty} \left(\frac{\lambda}{16\pi^2}\right)^{\ell}\mathcal{H}_{\ell} = \sum_{\ell = 0}^{\infty} \tilde{\lambda}^{\ell}\mathcal{H}_{\ell}
$$
where the terms $\mathcal{H}_{\ell}$ are expressed in terms of operators $P_{i,j}$ which switch the values of $a_{i}$ and $a_{j}$ in the operator.    For convenience, we adopt the notation of \cite{BMR} and introduce the structures
$$
\{a,b,c,...\} = \sum_{p = 1}^{L} \mathcal{P}_{p+a}\mathcal{P}_{p+b}\mathcal{P}_{p+c}\cdots
$$
where $\mathcal{P}_{\ell} = P_{\ell,\ell+1}$.  The Feynman diagrams contributing to this Hamiltonian naturally produce these structures.

The four-loop contribution to the dilatation operator acting on operators in the $SU(2)$ sector of length greater than four was found in \cite{BMR}.  We present it here together with the lower-loop contributions.

\begin{eqnarray*}
\mathcal{H}_{0} & = & \{\} \\
\\
\mathcal{H}_{1} & = & 2\{\} - 2\{1\} \\
\\
\mathcal{H}_{2} & = & -8\{\} + 12\{1\} - 2(\{1,2\} + \{2,1\}) \\
\\
\mathcal{H}_{3} & = & 60\{\} - 104\{1\} + 4\{1,3\} + 24(\{1,2\} + \{2,1\}) \\
&& -4i\epsilon_{2}\{1,3,2\} + 4i\epsilon_{2}\{2,1,3\} - 4(\{1,2,3\} + \{3,2,1\}) \\
\end{eqnarray*}
\begin{eqnarray*}
\mathcal{H}_{4} & = & (-560 - 4\beta_{2,3})\{\} \\
&& + (1072 + 12\beta_{2,3} + 8\epsilon_{3a})\{1\} \\
&& + (-84 - 6\beta_{2,3} - 4\epsilon_{3a})\{1,3\} \\
&& - 4\{1,4\} \\
&& + (-302 - 4\beta_{2,3} - 8\epsilon_{3a})(\{1,2\} + \{2,1\}) \\
&& + (4\beta_{2,3} + 4\epsilon_{3a} + 2i\epsilon_{3c} - 4i\epsilon_{3d})\{1,3,2\} \\
&&  + (4\beta_{2,3} + 4\epsilon_{3a} - 2i\epsilon_{3c} + 4i\epsilon_{3d})\{2,1,3\} \\
&& + (4 - 2i\epsilon_{3c})(\{1,2,4\} + \{1,4,3\}) \\
&& + (4 + 2i\epsilon_{3c})(\{1,3,4\} + \{2,1,4\}) \\
&& + (96 + 4\epsilon_{3a})(\{1,2,3\} + \{3,2,1\}) \\
&& + (-12 - 2\beta_{2,3} - 4\epsilon_{3a})\{2,1,3,2\} \\
&& + (18 + 4\epsilon_{3a})(\{1,3,2,4\} + \{2,1,4,3\} \\
&& + (-8 - 2\epsilon_{3a} - 2i\epsilon_{3b})(\{1,2,4,3\} + \{1,4,3,2\}) \\
&& + (-8 - 2\epsilon_{3a} + 2i\epsilon_{3b})(\{2,1,3,4\} + \{3,2,1,4\}) \\
&& - 10(\{1,2,3,4\} + \{4,3,2,1\})
\end{eqnarray*}

The values of the $\epsilon$'s above are not physical; they don't impact the spectrum and they vary depending on the gauge choice of the calculation and subtraction scheme.  The value of $\beta_{2,3}$ was found to be $\beta_{2,3} = 4\zeta[3]$ and comes from the dressing phase in the Bethe Ansatz diagonalization of the long-range Hamiltonian.  However, this Hamiltonian changes when it acts on an operator of finite size.  In the $SU(2)$ sector the first non-trivial such correction happens to $\mathcal{H}_{4}$ when acting on operators in the $SU(2)$ sector of length $L = 4$.  We will be looking at the mixing of operators $\mathcal{O}_1 = \mbox{Tr} X_1^2 X_2^2$ and $\mathcal{O}_2 = \mbox{Tr} X_1 X_2 X_1 X_2$.  The Hamiltonian in this sector is then a $2 \times 2$ matrix which we will express using the basis
$$
\mathcal{O}_1 = \left(\begin{array}{c} 1 \\ 0 \end{array}\right), \ \ \ \ \mathcal{O}_2 = \left(\begin{array}{c} 0 \\ 1 \end{array}\right).
$$
One combination of these two operators is the BPS operator
$$
\mathcal{O}_{\mbox{BPS}} = 4\mathcal{O}_{1} + 2\mathcal{O}_{2}.
$$
The other (exact) eigenstate of the dilatation operator is the Konishi operator \cite{ryzhov}
$$
\mathcal{O}_{K} = \mathcal{O}_{1} - \mathcal{O}_{2}.
$$
The dilatation operator in this sector can, in fact, be written in terms of a single $2 \times 2$ matrix $\mathcal{M}$ and an over-all constant that is an expansion in the 't Hooft coupling (and gives the loop by loop corrections to the dimension of the Konishi operator.)  Specifically, we can write
$$
\mathcal{H}^{K}_4 = C_{4}\mathcal{M} = C_{4}\left(\begin{array}{cc} 2 & -4 \\ -2 & 4 \end{array}\right)
$$
where $\mathcal{H}^{K}_4$ is the four-loop term in the expansion of the dilatation operator, acting on the length 4 sector mixing $\mathcal{O}_1$ and $\mathcal{O}_2$.  (Note that $\mathcal{M}$ vanishes when acting on the BPS state.)  It will sometimes be convenient to express the matrix $\mathcal{M}$ as a sum of two terms
$$
\mathcal{M} = 4I - M, \ \ \ \ \ I = \left(\begin{array}{cc} 1 & 0 \\ 0 & 1 \end{array}\right), \ \ \ \ \ M = \left(\begin{array}{cc} 2 & 4 \\ 2 & 0 \end{array}\right)
$$

The change to $\mathcal{H}_{4}$ can be expressed in terms of two sets of diagrams: those that appear in the calculation when $L > 4$ which disappear for $L = 4$ we will include in a term $\tilde{H}_{u}$, those that don't exist for $L > 4$ but appear for $L = 4$ we will include in a term $\tilde{H}_{w}$.  We are then looking for
$$
\mathcal{H}^{K}_4 = \mathcal{H}_{4} - \tilde{H}_{u} + \tilde{H}_{w}.
$$

\begin{figure}
\begin{center}
\resizebox{7in}{!}{\includegraphics{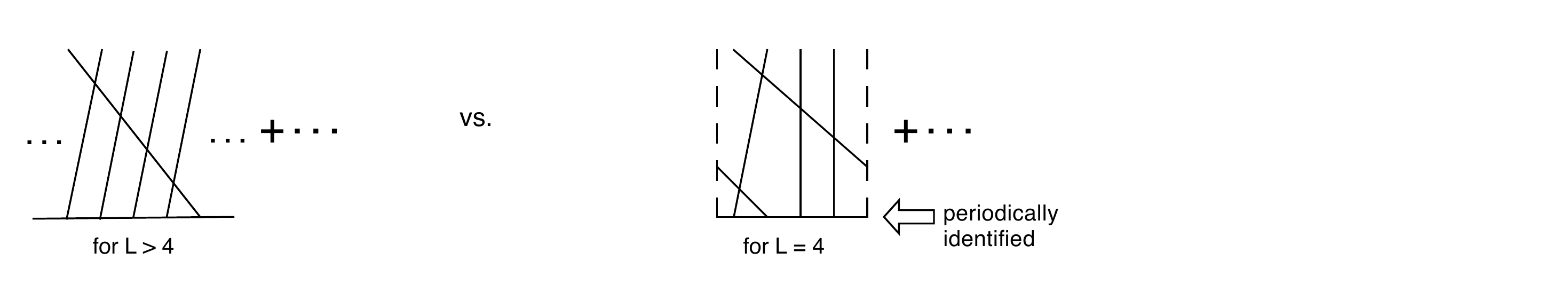}}
\end{center}
\caption{\label{fourloop} Diagrams that exist at four loops for operators of length $L > 4$ and those for length $L = 4$.}
\end{figure}

$\tilde{H}_{u}$ consists of four loop diagrams that are of ``maximal length''; that is, they involve the maximum number of fields in the operator.  $\tilde{H}_{w}$ consists of four loop ``wrapping diagrams'', diagrams that explicitly wrap around the operator and would not be possible for longer operators at this loop level.  Examples are shown in Figure \ref{fourloop}.

To simplify matters, we will actually not compute $\tilde{H}_{w}$ and $\tilde{H}_{u}$ but instead work with objects $H_w$ and $H_u$ which will be defined so that they separately vanish on BPS states.  This will allow us to ignore any diagram which only contributes to the identity, and at the end fix the coefficient of the identity by requiring the BPS condition.  In the next three sections we will give a description of all diagrams contributing to $H_{u}$ and $H_{w}$, calculate $H_u$, calculate $H_w$, and finally give our result for the anomalous dimension of the Konishi operator to four loops.

\section{Wrapped and Unwrapped: The diagrams
necessary}\label{notation}
\subsection{Our Diagram Notation: The Unwrapped Case}

First let us consider those diagrams which are present in the length
5 operator case at order $\lambda^4$, but are too wide to fit on a
length 4 operator.  These unwrapped maximal length diagrams will be
further discussed in Section 4.  The diagrams at issue are those
which contain non-separable interaction over five sites (no six site
interaction is possible at order $\lambda^4$).

\begin{figure}
\begin{center}
\resizebox{.5in}{!}{\includegraphics{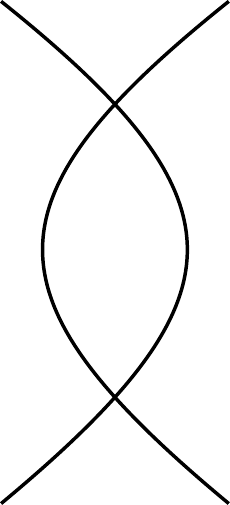}}
\end{center}
\caption{\label{scalarloop} An example of a diagram which is \emph{not} maximal length.}
\end{figure}

\begin{figure}
\begin{center}
\resizebox{.5in}{!}{\includegraphics{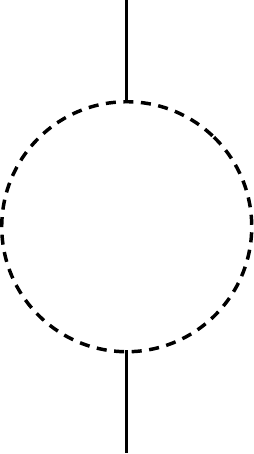}}
\end{center}
\caption{\label{fermionloop} A fermion loop in an unwrapped diagram cannot lead to a diagram of maximal length.}
\end{figure}

All five-site diagrams at order $\lambda^4$ can only be created by
having one single interaction between each neighboring pair of
scalars.  For example interactions as in Figure \ref{scalarloop} are
order $\lambda^2$ by themselves.  As such, any diagram of order
$\lambda^4$ which includes this piece can be no wider than four
sites. Similarly, no fermion interactions can be part of this
maximal width group. For example, consider Figure \ref{fermionloop}.
It shows an interaction which has order $\lambda$ by itself; at
order $\lambda^4$, any diagram containing this interaction can only
be four sites wide.

Thus, we can label all 5-site interactions by listing whether a
scalar or gluon interaction occurs between each neighboring pair of
sites.  We use X to represent the 4-scalar interaction, and G to
represent gluon exchange.  Thus $XXXX$ represents a 5-site diagram
which consists of entirely 4-scalar interactions, while $XGGG$ means
that a scalar interaction occurred between the first two sites, but
all other interactions are gluon exchange.

\begin{figure}
\begin{center}
\resizebox{3in}{!}{\includegraphics{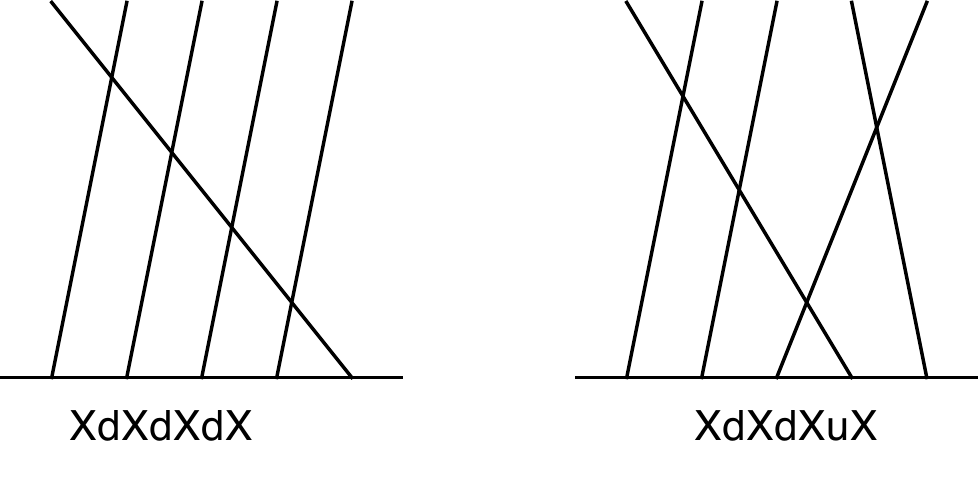}}
\end{center}
\caption{\label{XXXX} Two maximal length diagrams involving four scalar interactions in different orders.}
\end{figure}

Of course this doesn't completely specify the diagram under
consideration; for example, consider the diagrams in Figure
\ref{XXXX}.  Both of these diagrams are five site diagrams with a
single scalar interaction between each pair of neighboring sites;
however, they will give different integrals as well as different
tensor structures.  We need a way of distinguishing these
interactions, while also ensuring that we count all possible
diagrams.  It can be shown that only the vertical ordering of nearest
neighboring interactions matters;  thus if we insert $u$ when the
right interaction of a pair occurs above the left, and $d$ when it
occurs below, we will completely distinguish between diagrams.  In
this notation, the left diagram of Figure \ref{XXXX} is labeled
$XdXdXdX$, while the right one is $XdXdXuX$.

\begin{figure}
\begin{center}
\resizebox{1.25in}{!}{\includegraphics{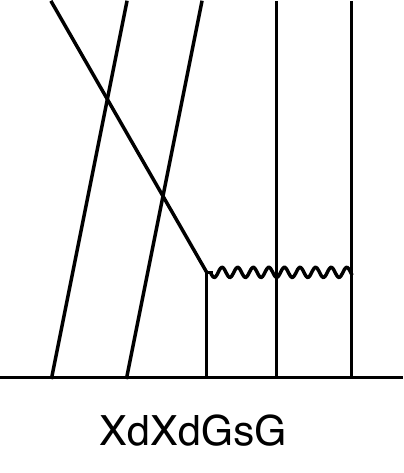}}
\end{center}
\caption{\label{XdXdGsG} An example of a diagram with a two-gluon/two-scalar vertex.}
\end{figure}

When we include gluon interactions we must also allow for the
two-gluon/two-scalar vertex, which means that neighboring gluon
interactions can occur at the same vertical position; we call this $s$.  As an
example, the diagram in Figure \ref{XdXdGsG} is written as $XdXdGsG$.

Thus the rules for producing the five-site unwrapped diagrams at
$\lambda^4$ are:

\begin{enumerate}
\item
Choose either $X$ or $G$ for each of the four interactions between
neighboring sites, for example $XXGG$.
\item
Between pairs of $XX$ or $XG$, decide the vertical ordering of the
interaction; insert $u$ or $d$ accordingly.  Between two gluon
interactions, $s$ is also possible.  Thus in our example, for
diagrams of type $XXGG$, there are 12 possibilities.
\end{enumerate}

Following these rules will produce all diagrams which are present
for length 5 operators, but not for length 4 operators, at loop
order $\lambda^4$.  These rules can clearly be extended to the
general length $L$ case.

\subsection{Wrapped diagrams}

\begin{figure}
\begin{center}
\resizebox{3.25in}{!}{\includegraphics{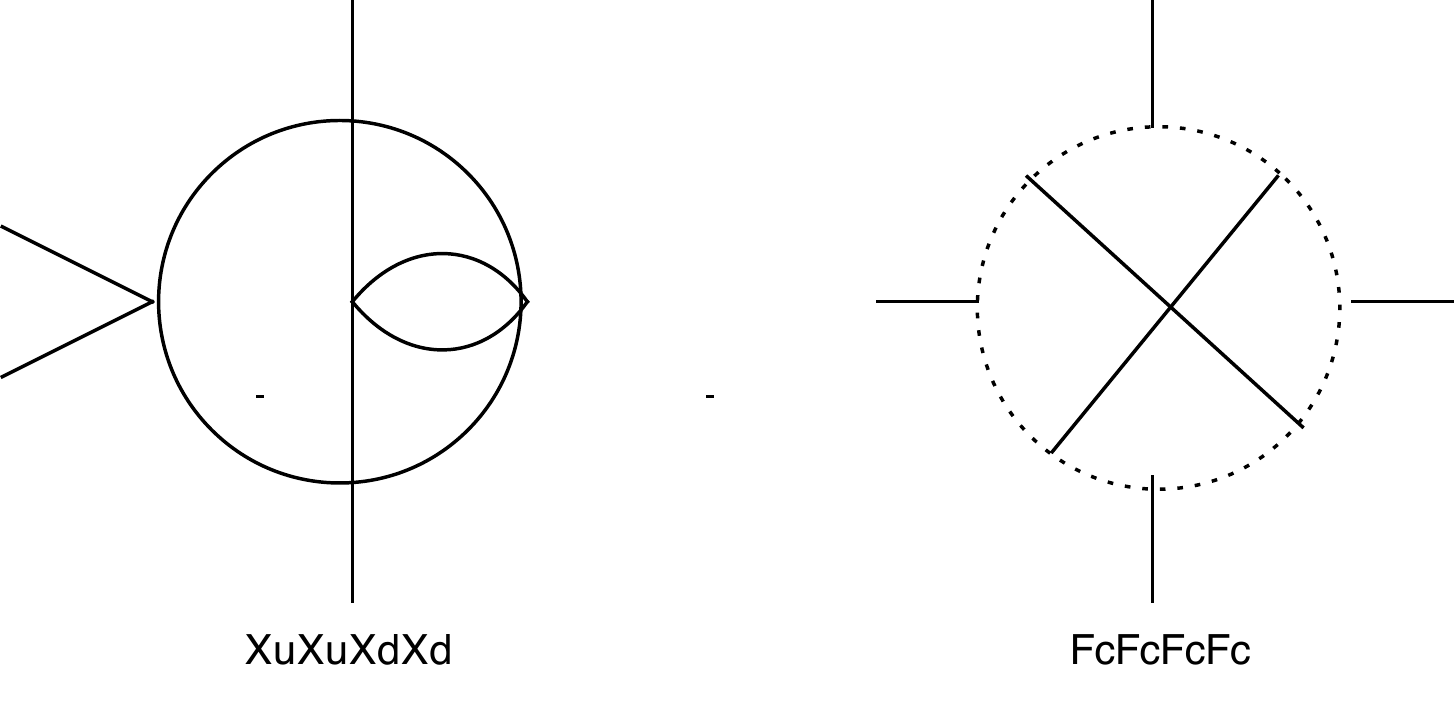}}
\end{center}
\caption{\label{wrappeddiagrams} Two examples of wrapped diagrams.}
\end{figure}

Now let us consider diagrams which are present at order $\lambda^4$
for the length 4 operators, but not for length 5.  These diagrams
are those which ``wrap around'' a 4 site operator, and will be
further discussed in Section 5. Two such diagrams, drawn with the
operator inserted at the center, are shown in Figure
\ref{wrappeddiagrams}. All diagrams of this variety either involve
only fermion interactions, as in the right hand diagram of Figure
\ref{wrappeddiagrams}, or none, as in the left-hand diagram.

First let us consider the non-fermion diagrams.  Again these
diagrams consist of a single scalar or gluon interaction between
each pair of neighboring sites, and the vertical ordering only matters
between neighboring interactions.  Thus, in the example of an $XXXX$
type diagram, we must now decide which interaction occurs first
between all four pairs.  There are therefore sixteen diagrams of
this type; two examples are $XuXuXuXu$ or $XuXuXuXd$.  Of course if
we are considering a diagram of type $GXXG$ we can also include $s$
as a choice between the two $G$ interactions.

\begin{figure}
\begin{center}
\resizebox{2in}{!}{\includegraphics{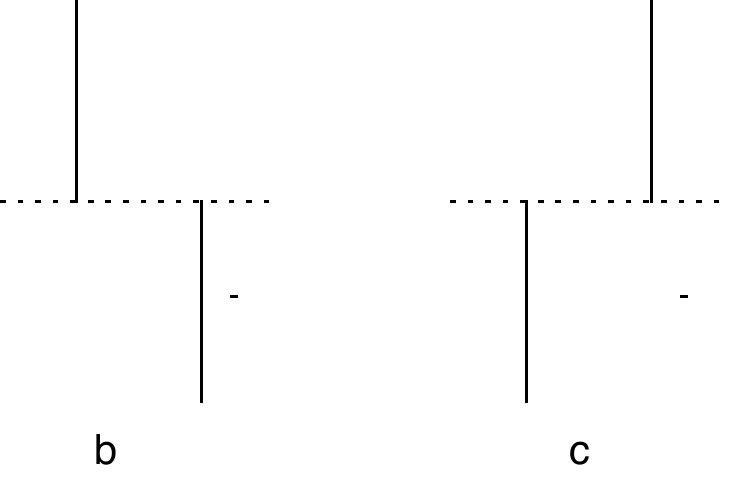}}
\end{center}
\caption{\label{bc} The building blocks for diagrams involving fermions.}
\end{figure}

Now let us consider the fermion diagrams.  We find that a fermion loop is possible in the wrapping case if it wraps all the way around the operator.  The diagrams are always built out of the two blocks shown in Figure \ref{bc}, (three incoming $SU(2)$ scalars attaching to the fermion loop right next to each other can be shown to vanish because of the Clebsch-Gordon matrices.)  If we were to allow for operators outside this sector, this would not be true.  Within this restriction however, we have a choice of either $b$ or $c$ for each of the four sites, leading to 16 diagrams.  Figure \ref{wrappeddiagrams} showed the diagram $FcFcFcFc$.  Again, these rules can clearly be extended to the length $L$ case.  

\subsection{Diagram Symmetries}

We can reduce the number of separate diagrams we have to compute by
noticing two types of symmetries in the diagrams.  First, we can
reflect the diagram, resulting in the same integral.  For
example, $XuXdXdX$  represents the same integral as $XuXuXdX$; however they are two different diagrams and we must include each (including their different tensor structures). In
general, this reflection symmetry means we need only calculate one integral of each such pair, but must multiply by the sum of the tensor structures.

To find the reflection of a general scalar/gluon diagram, write the diagram in the opposite order while also switching $u$ for $d$ and $d$ for $u$ ($s$ does not switch).  Thus, $XuGuXdX$ will produce
the same integral as $XuXdGdX$.  This switching from $u$ to $d$ is logical because $u$ indicates that the right hand interaction occurs above the one on the left, whereas $d$ indicates the reverse. Similarly, fermion diagrams reflect by switching $b$ for $c$ and vice versa. Note that some diagrams, for example $XuGsGdX$, are their own reflection.  For these diagrams, we want to make sure to include only one copy of their tensor structure.

Wrapped diagrams exhibit one additional degree of symmetry, in addition to the reflection equivalence.  Because the trace is cyclic, these diagrams exhibit rotational symmetry.  As an example, $XuXuXuXd$ and $XdXuXuXu$ give the same integral result.  However, as they arise from different contraction patterns, we must sum over both their tensor structure contributions (as well as all other rotations and reflections, a total of eight diagrams).  These two symmetry forms help reduce the number of different integrals to a manageable level.

This notation system additionally allows us to reduce calculating
the number of diagrams which contribute to the first order wrapping
effect (at loop order $L$, the difference between $L+1$ and $L$
sites) to a combinatorial problem.  For the case of order 4 which we
consider in this paper, there are 187 diagrams present in the 5-site
case which cannot be drawn for 4 sites, and 449 wrapped diagrams
which are only present for 4 sites, for a total of 636 diagrams. Of
course, given the symmetries mentioned in the previous section, we
only have to calculate separate integrals for a fraction of these
diagrams.  We further reduce the number of integrals to calculate by
considering only those diagrams whose tensor structure is nontrivial
on the $SU(2)$ subsector.

\section{Unwrapped Maximal Length Diagrams}

We now wish to use the unwrapped maximal length diagrams to compute $H_u$.  A complete list of the counterterms from these diagrams is included in the appendix.\footnote{Where diagrams are related by symmetry, such as the counterterms for $XuXuXuX$ and $XdXdXdX$, we have included only one copy in the appendix.}  We must first determine what diagrams contribute to what $\{a,b,c,...\}$ structures.  Consider a general unwrapped diagram with four interactions, either gluon or scalar, and three ordering labels $u,d,s$.  Each scalar interaction involves a factor of $2P_{i,i+1} - I_{i,I+1}$ acting on the sites between which the interaction sits.  This should be clear from, for example, \cite{MZ}.  Because a gluon exchange does not affect the R-charge of the scalars, each gluon interaction carries a factor of $I_{i,i+1}$.  In addition, these factors need to be correctly ordered, with the ordering given by the $u,d,s$ structure of the diagram.  The ordering proceeds from top to bottom.  Thus, for example, the diagram $XuGuXdX$ has the structure
$$
XuGuXdX: \ \sum_{i }(2P_{i+2,i+3} - I_{i+2,i+3})(2P_{i+3,i+4} - I_{i+3,i+4})I_{i+1,i+2}(2P_{i,i+1} - I_{i,i+1})
$$
where we notice that the ambiguity in ordering of the last three terms is ok because $[P_{i,i+1}, P_{i+3,i+4}] = 0$.  Now it becomes clear that for diagrams with gluons in them we can make sets of diagrams which will lead to the same tensor structure.  The diagrams $XuGuXdX$, $XuGdXdX$, $XdGuXdX$ and $XdGdXdX$ all have the same structure because the orderings adjacent to the gluon, which only carries a factor of the identity, are irrelevant.  The four $XXXX$ diagrams each stands alone, but all others fall into sets.  And here the power of supersymmetry becomes evident, because we find that the sum of counterterms in each of these sets other than the $XXXX$ diagrams leaves no simple $1/\epsilon$ pole, and therefore does not contribute to $H_{u}$.  It appears that \emph{the only} unwrapped diagrams one need consider are the $XXXX$ diagrams.  These lead to the structures
\begin{eqnarray*}
XuXuXuX & : &  \ 8\{1,2,3,4\} - 4\{1,2,4\} - 4\{1,3,4\} - 8\{1,2,3\} \\
&& + 6\{1,2\} + 4\{1,3\} + 2\{1,4\} - 4\{1\} + (1/2)\{\}
\end{eqnarray*}
\begin{eqnarray*}
XuXuXdX & : & \ 8\{3,2,1,4\} - 4\{3,2,1\} - 4\{2,1,4\} - 4\{3,1,4\} - 4\{2,1,3\} \\
&& + 6\{1,2\} + 4\{1,3\} + 2\{1,4\} - 4\{1\} + (1/2)\{\}
\end{eqnarray*}
\begin{eqnarray*}
XuXdXuX& : & \ 8\{2,4,1,3\} - 4\{2,1,4\} - 4\{4,1,3\} - 4\{2,1,3\} - 4\{1,3,2\} \\
&& + 6\{1,2\} + 4\{1,3\} + 2\{1,4\} - 4\{1\} + (1/2)\{\}
\end{eqnarray*}
\begin{eqnarray*}
XdXuXuX & : & \ 8\{1,4,3,2\} - 4\{1,4,3\} - 4\{1,4,2\} - 4\{1,3,2\} - 4\{3,2,1\} \\
&& + 6\{1,2\} + 4\{1,3\} + 2\{1,4\} - 4\{1\} + (1/2)\{\}
\end{eqnarray*}
where we have fixed the normalization so that the residues of the terms in the appendix can be used directly, and where the other four scalars diagrams, those that are related by reflection symmetry to the ones above, have structures determined by the rule $1 \leftrightarrow 4$, $2 \leftrightarrow 3$.  Without yet inputting values, we then find that
\begin{eqnarray*}
H_u & = & (8rXuXuXuX)\big(\{1,2,3,4\} + \{4,3,2,1\}\big) \\
&& + (8rXuXuXdX)\big(\{3,2,1,4\} + \{2,1,3,4,\}\big) \\
&& + (8rXdXuXuX)\big(\{1,4,3,2\} + \{1,2,4,3\}\big) \\
&& + (8rXuXdXuX)\big(\{2,1,4,3\} + \{1,3,2,4\}\big) \\
&& + (-8rXuXuXuX - 4rXuXuXdX - 4rXdXuXuX)\big(\{1,2,3\} + \{3,2,1\}\big) \\
&& + (-4rXuXuXuX - 8rXuXuXdX - 4rXuXdXuX)\big(\{1,3,4\} + \{2,1,4\}\big) \\
&& + (-4rXuXuXuX - 4rXuXdXuX - 8rXdXuXuX)\big(\{1,2,4\} + \{1,4,3\}\big) \\
&& + 2(-4rXuXuXdX - 4rXuXdXuX)\{2,1,3\} \\
&& + 2(-4rXuXdXuX - 4rXdXuXuX)\{1,3,2\} \\
&& + (6rXuXuXuX + 6rXuXuXdX + 6rXuXdXuX + 6rXdXuXuX)\big(\{1,2\} + \{2,1\} \\
&& + 2(2rXuXuXuX + 2rXuXuXdX + 2rXuXdXuX + 2rXdXuXuX)\{1,4\} \\
&& + 2(4rXuXuXuX + 4rXuXuXdX + 4rXuXdXuX + 4rXdXuXuX)\{1,3\} \\
&& + 2(-4rXuXuXuX - 4rXuXuXdX - 4rXuXdXuX - 4rXdXuXuX)\{1\}. \\
\end{eqnarray*}

Now, the prefix ``r'' in front of the diagram means that we should take the residue of the counterterm, and that the coefficient of $\{\}$ is chosen to be zero so that the entire $H_u$ vanishes on a BPS state.  We expect that all the structures in $H_u$ which are of maximal length (which involve ``4'') should match those in $\mathcal{H}_4$.  Comparison with the values in the appendix
$$
\begin{array}{cc}
rXuXuXuX = -\frac{5}{4} & \ \ \ \ \ \ 
rXuXuXdX = \frac{1}{3} \\ \\
rXdXuXuX = -\frac{1}{3} & \ \ \ \ \ \ 
rXuXdXuX = \frac{1}{4} \\
\end{array}
$$
shows agreement with the assignments
$$
\epsilon_{3a} = -4, \ \ \ \ \ \ i\epsilon_{3b} = \frac{4}{3}, \ \ \ \ \ \ i\epsilon_{3c} = \frac{-4}{3}.
$$
This yields
\begin{eqnarray*}
\mathcal{H}_{4} - H_u & = & (-560 - 4\beta_{2,3})\{\} \\
&& + (1064 + 12\beta_{2,3} + 8\epsilon_{3a})\{1\} \\
&& + (-76 - 6\beta_{2,3} - 4\epsilon_{3a})\{1,3\} \\
&& + (-296 - 4\beta_{2,3} - 8\epsilon_{3a})\big(\{1,2\} + \{2,1\}\big) \\
&& + (10 + 4\beta_{2,3} + 6\epsilon_{3a} - 2i\epsilon_{3b} + 2i\epsilon_{3c} - 4i\epsilon_{3d})\{1,3,2\} \\
&& + (10 + 4\beta_{2,3} + 6\epsilon_{3a} + 2i\epsilon_{3b} - 2i\epsilon_{3c} + 4i\epsilon_{3d})\{2,1,3\} \\
&& + (78 + 2\epsilon_{3a})\big(\{1,2,3\} + \{3,2,1\}\big) \\
&& + (-12 - 2\beta_{2,3} - 4\epsilon_{3a})\{2,1,3,2\}
\end{eqnarray*}

We now want to assume that this acts on an operator of length $L = 4$ and express it in terms of the matrix $\mathcal{M}$.  Table \ref{tensors} shows how to relate the $\{a,b,c,...\}$ structures to the matrices $M$ and $I$.

\begin{table}
\begin{center}
\begin{tabular}{c|c}
$4I = \left(\begin{array}{cc} 4 & 0 \\ 0 & 4 \end{array}\right)$ & $M = \left(\begin{array}{cc} 2 & 4 \\ 2 & 0 \end{array}\right)$ \\
& \\
\hline
& \\
$\{\}$ & $\{1\}$ \\
& \\
$\{1,3\}$, $\{2,1,3,2\}$ & $\{1,2\}$, $\{2,1\}$ \\
& \\
$\{1,2,3\}$, $\{3,2,1\}$ & $\{2,1,3\}$, $\{1,3,2\}$ \\
\\
\end{tabular}
\end{center}
\caption{\label{tensors} A table showing how to relate tensor structures to the matrices $M$ and $I$, assuming the length of the operator is $L = 4$.}
\end{table}

Finally, the result for this section is
$$
\mathcal{H}_{4} - H_u = 4(123 + \epsilon_{3a} + 3\beta_{2,3})(M - 4I) = -4(119 + 12\zeta[3])\mathcal{M}.
$$

Note that this result is gauge dependent.  The final result $\mathcal{H}_{4} - H_u + H_w$ should not be, but we will not see this in an obvious way as it is less convenient to keep track of the gauge dependent parameters in the wrapping part of the calculation.

These diagrams, along with the wrapping ones, were each checked in at least one of three ways.  Patterns were found in the counterterms, for example it was discovered that replacing any top gluon exchange interaction with a scalar vertex, or vice versa, has no effect on the counterterm.  Some diagrams were checked by noting that they conformed to this pattern.  Furthermore, some diagrams can be computed in more than one way, for example with more than one choice of external momenta.  These diagrams were computed both ways and the results compared.  Finally, some diagrams were checked by having the two authors separately compute the same diagram, and compare results.

\section{Wrapped Diagrams and the Final Answer}

We now wish to calculate $H_w$ from the wrapping diagram results listed in the appendix.  We again want to determine what the tensor structure associated to each diagram calculated is.  Here we group the 449 diagrams accounted for in section three into sets that are associated to the same integral and work out what the tensor structure for the sum is.  For example, we find
\begin{eqnarray*}
XuXuXuXd + XuXuXdXu + XuXdXuXu + XdXuXuXu && \hspace{2in} \\
+ XdXdXdXu + XdXdXuXd + XdXuXdXd + XuXdXdXd & \rightarrow & 8M - 28I
\end{eqnarray*}

Note that this includes all diagrams related by either rotation or reflection.  We ignore diagrams of the type $GGGG$ and of the type $XXGG$ because these diagrams contribute only to the identity.\footnote{The latter fact is a result of the the non-trivial tensor structures canceling out, by $2\{1,2\} - 2\{1\} + (1/2)\{\} = 2I$.}  The tensor structures of the fermion loop diagrams must be worked out by referring to the Clebsch-Gordon matrices present at each vertex.  Roughly, every time there is a $cb$ structure, we pick up a factor of $(P_{i,i+1} - I_{i,i+1})$ because the two adjacent incoming scalar fields must not have the same R-charge for the diagram not to vanish.  It turns out that $FcFcFcFc$ and $FbFbFbFb$ therefore only contribute to the identity and can be ignored.  This leaves 45 separate integrals which must be computed; the combination of them which is proportional to $M$ in $H_{w}$ is
$$
-4rXuXuXuXu + 8rXuXuXuXd - 4rXuXuXdXd - 6rXuXdXuXd - 2rXuXuXuGu
$$
$$
- 2rXuXuXdGu - 2rXuXuXuGd - 2rXuXuXdGd + 6rXuXdXuGu + 3rXuXdXdGu
$$
$$
+ 3rXuXdXuGd + 6rXdXuXuGu + 3rXdXuXdGu + 3rXdXuXuGd - 2rXuGuXuGu
$$
$$
- 4rXuGuXuGd - 4rXuGuXdGu - 2rXuGuXdGd - 2rXuGdXdGu - rXuGdXuGd
$$
$$
- rXdGuXdGu + 2rXuGuGuGu + 2rXuGuGuGd + 2rXuGuGdGu + 2rXuGdGuGu
$$
$$
+ 2rXdGuGuGu + 2rXuGdGdGu + rXuGuGdGd + rXdGdGuGu + rXuGdGuGd
$$
$$
+ rXdGuGdGu + 2rXuGuGsGu + 2rXuGuGsGd + 2rXuGdGsGu + 2rXdGuGsGu
$$
$$
+ 2rXuGdGsGd + 2rXdGdGsGu + 2rXdGuGsGd + 2rXdGdGsGd + 2rXuGsGsGu
$$
$$
+ rXuGsGsGd + rXdGsGsGu - 8rFcFcFbFb - 8rFcFbFcFb - 16rFcFcFcFb
$$
When doing this analysis it is important to count each separate diagram only once.  For example, naively one might expect that the coefficients in front of $XdGuXdGu$ and $XuGuXdGu$ should be the same because changing the ordering around a gluon does not affect the tensor structure of a diagram.  However, there are eight distinct diagrams which contribute to the tensor structure represented by $XuGuXdGu$ while there are only two contributing to the tensor structure represented by $XdGuXdGu$: $XdGuXdGu$ and $GuXdGuXd$.  This leads to the factor of 4 difference between the coefficients.   Sadly, the cancellations that occurred in the unwrapped diagrams do not seem to be present in this case, so the answer cannot be written in terms of just the scalar diagrams.  Using the results from the appendix, we find
$$
H_w = \left(\frac{83}{2} + 20\zeta[3] - 140\zeta[5]\right)\mathcal{M}
$$
which gives a final answer of
$$
\mathcal{H}^{K}_{4} = -\left(\frac{869}{2} + 28\zeta[3] + 140\zeta[5]\right)\mathcal{M}
$$
or
$$
\Delta^{K} = 4 + 12\tilde{\lambda} - 48\tilde{\lambda}^2 + 336\tilde{\lambda}^3 - (2607 + 28\zeta[3] + 140\zeta[5])\tilde{\lambda}^4 + \cdots
$$
for the dimension of the Konishi operator, expanded out to four loops.  (The three loop result for the Konishi operator was found in \cite{threeloops}.)

\section{Future Directions}

The most important direction to take this research in at this point is towards a generalization for the $L$-loop correction to the anomalous dimensions of operators with length $L$.  This would represent the first finite size correction to the Beisert-Eden-Staudacher proposal for the anomalous dimensions of long operators.  We could then compare this correction to recent proposals for finite size corrections on the string theory side of the correspondence \cite{Janik}.  Furthermore, we could examine the form of the correction to see if it is consistent with integrability.  Although we do not have this general result, we do have some thoughts on the matter.

An important question to ask is what the size of the first finite size correction should be.  We know that perturbatively in the gauge theory the correction will be suppressed by $\lambda^{L}$ where we assume that $\lambda$ is small.  This is clear because this is the loop order at which the wrapping effect first appears.  However, we also expect finite size corrections to be small on the string theory side, where $\lambda$ is large.  There must therefore be some additional suppression.  One idea for the origin of this suppression comes from the relatively small number of diagrams that contribute to the wrapping effect, as opposed to those that contribute to the anomalous dimension as a whole.  Indeed, this limitation was the inspiration for this project because it meant that the number of diagrams necessary to calculate the anomalous dimension of the Konishi operator to four loops was much smaller than if we didn't already have the result for long operators.  A naive counting of just the scalar diagrams suggests a suppression that is non-perturbative in $1/L$, roughly of the form $\frac{(L!)^2}{(3L)!}$.

Another significant issue to address is what the degree of trancendentality of the finite size correction should be.   One important result of this calculation is the fact that the degree of trancendentality for the anomalous dimension of the Konishi operator at four loops is $5$, while that for a long operator is $3$.  Although we cannot know for certain what the degree will be for arbitrary loop and length $L$, we can examine specific diagrams which generalize easily.  For example, the calculation will always involve a diagram of the form $XuXuXuXu...$ for an arbitrary number of scalar interactions.  This is a particularly convenient diagram to consider because it has not associated daughter diagrams.  For us, the counterterm $cXuXuXuXu$ had a residue of $-5\zeta[5]$, which reflected the maximum degree of trancendentality of the entire calculation.  For general $L$, the equivalent diagram would involve a $\zeta[2L - 3]$.  If this can be taken as an indication of the trancendentality of the finite size correction, then it continues to lead that of the long operator at the same loop order.

Finally, we note that the calculation presented here relied on the explicit knowledge of $\mathcal{H}_{4}$.  For higher loops, we don't have an expression for the dilatation operator acting on long operators.  What we have is the spectrum, given in terms of a Bethe Ansatz.  However, it should be possible to set up a version of a standard, quantum mechanical perturbation analysis, using the Bethe Ansatz to give the diagonalization of the unperturbed Hamiltonian, and thinking of $H_u$ and $H_w$ as perturbations.  The complication here is that when we change the size of the spin chain, we are actually changing the Hilbert space.  However, it still should be possible to perform the calculation using expectation values.  This possibility, as well as the above thoughts about correction size and trancendentality, will be further explored in a forthcoming paper \cite{KM2}.

\section{Acknowledgments}

We would like to thank Ilarion Melnikov, Oleg Lunin, Matthias Staudacher, Tristan McLoughlin, Petr Horava, and Nikolas Beisert for many illuminating discussions.  We would especially like to thank Radu Roiban for his kind assistance with various technical issues as well as discussions.   The work of N.M. was partially supported by Department of Energy grant 580093 and partially supported by the Enrico Fermi Institute endowment 750573.  In addition, part of the work for this paper was done while N.M. was a visitor to the Berkeley CTP, and part while a visitor to the Cambridge INI.  The work of C.K. was supported by an NSF Graduate Research Fellowship.

\appendix

\section{Diagram Calculations}

\subsection{Basic Diagram Calculation Procedure}

The terms $\mathcal{H}_{\ell}$ in the dilatation operator are calculated by requiring a two-point function
$$
\langle X^{a_1}X^{a_2}\cdots X^{a_{L}}\mathcal{O}^{I}\rangle
$$
to be finite under renormalization
$$
X \rightarrow Z_{X}X, \ \ \ \ \ \mathcal{O}^{I} \rightarrow (Z_{\mathcal{O}})^{I}_{J}\mathcal{O}^{J}.
$$

If we know, $Z_{\mathcal{O}}$, then we can determine the dilation operator by the relation
$$
\mathcal{H} - \mathcal{H}_{0} = \frac{\epsilon}{2}Z^{-1}_{\mathcal{O}}\frac{d Z_{\mathcal{O}}}{d\ln \lambda}
$$
where $\epsilon$ is the dimensional regularization parameter such that $d = 4 - 2\epsilon$.  $Z_{\mathcal{O}}$ consists of an expansion in $\lambda$ and contains the counterterms associated to diagrams in the correlator at each loop order.  Note that these counterterms can have of divergences in $\epsilon$ that go like $1/\epsilon^{n}$ for general positive integer $n$.  However, these must be organized so that $\ln Z_{\mathcal{O}}$ has only $1/\epsilon$ poles; this insures that the final answer is finite and independent of $\epsilon$.  However, the $1/\epsilon$ poles in $\ln Z_{\mathcal{O}}$ can only come from the $1/\epsilon$ poles in $Z_{\mathcal{O}}$.  We can therefore ignore all but the residues of the counterterms, and use them directly to calculate $\mathcal{H}_{\ell}$.

The Feynman rules necessary for the calculation of these diagrams have been presented in many papers, for example in \cite{GMR}.  Of importance to us are the complex scalar propagator, which carries a factor of $\frac{1}{p^2}$, the gluon propagator, which carries a factor of $\frac{1}{2p^2}$, the vertex of four complex scalars, which carries a factor of $\frac{1}{2}$, and the vertex of two complex scalars and a gluon, which carries a factor of $1$ in addition to the dependence on the momenta of the scalars.  In addition we need the fermion propagator $\frac{\sigma\cdot p}{2p^2}$ and the fermion-fermion-scalar vertex, which carries a factor of the Clebsch-Gordon matrices relating the $\mathbf{4}$ and $\mathbf{6}$ representations of $SU(4)$.  For an explicit expression of these see for example \cite{EM}.  For convenience, all of these factors of $2$ and $1/2$ have been subsumed into the tensor structures of the diagrams; the integrals presented in the final sections of the appendices assume all propagators and vertices have factors of $1$.

\begin{figure}
\begin{center}
\resizebox{2in}{!}{\includegraphics{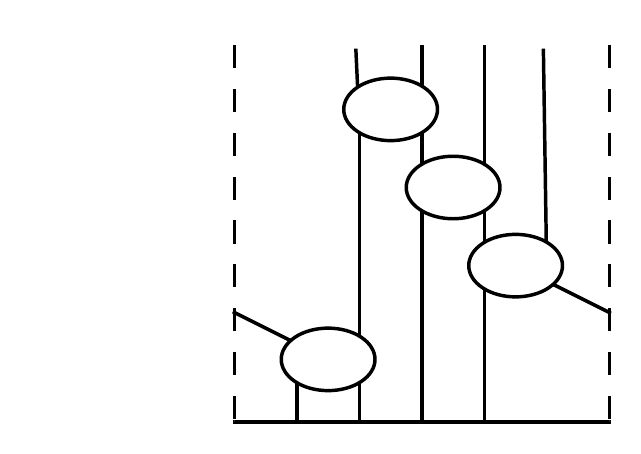}}
\end{center}
\caption{\label{generalwrapped} The basic form of a wrapped diagram.}
\end{figure}

Diagrams in the correlator of an operator with external fields take the general form shown in Figure \ref{generalwrapped} for the wrapping diagrams, (where the vertical ordering is arbitrary,) and a similar form for the unwrapped diagrams.

\begin{figure}
\begin{center}
\resizebox{5in}{!}{\includegraphics{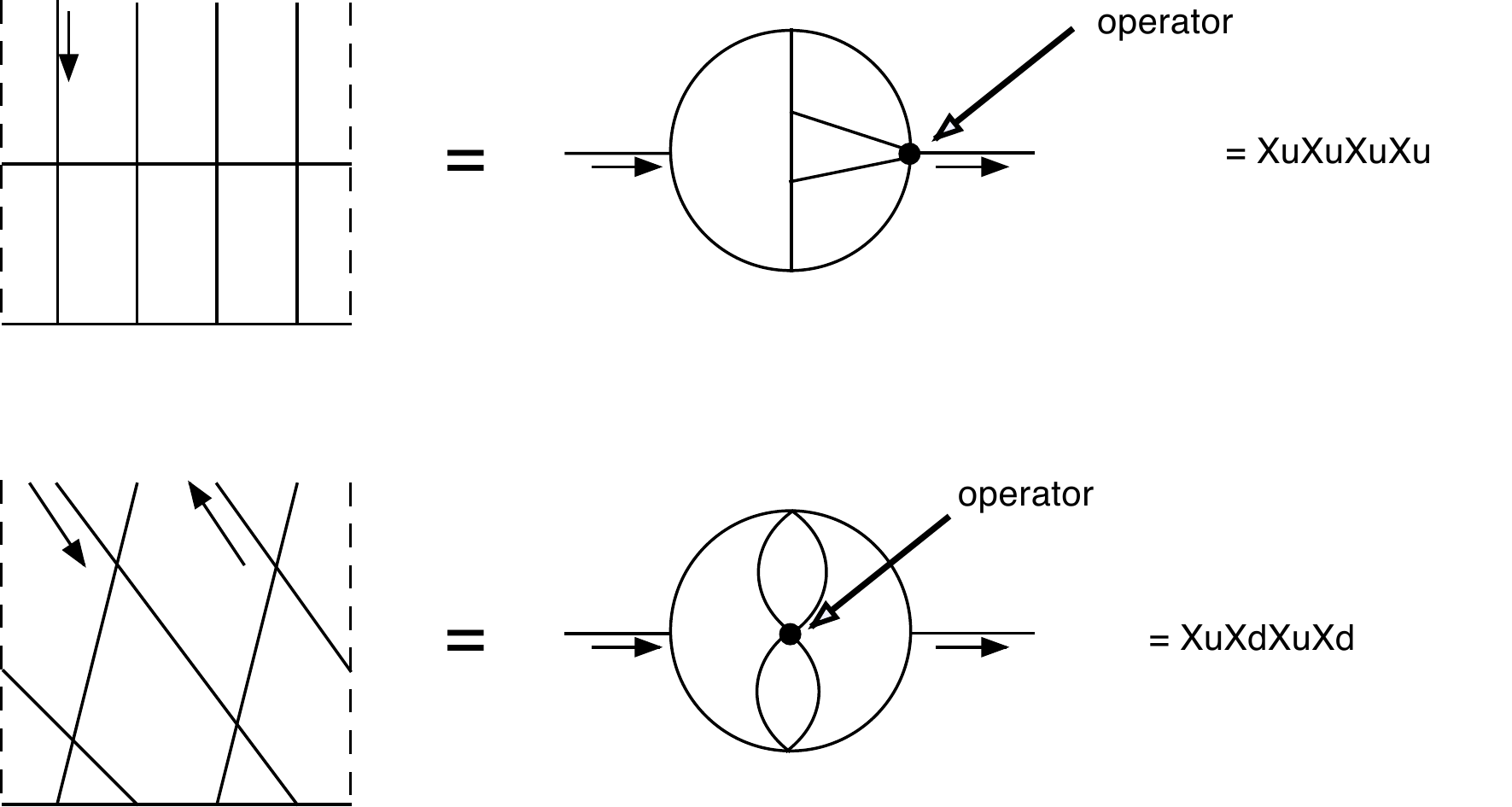}}
\end{center}
\caption{\label{momenta} Incoming momenta choices for the diagrams $XuXdXuXd$ and $XuXuXuXu$.  Note that both can be expressed as the correction to a propagator.}
\end{figure}

The counterterm associated with a given diagram (added to the necessary lower-loop counterterm diagrams, as explained in the next section) is independent of the incoming momenta, with the restriction that they be chosen so as not to introduce any IR divergences into the diagram.  (The renormalization of the operator ought to be dependent only on UV behavior.)  Depending on the vertical ordering of interactions in the diagram, the simplest possible choice of momenta varies.  We find that in order to avoid IR divergences it is necessary to have momentum flow into any ``top'' interaction.  If there are two gluon interactions at the top tied together with an "s" ordering there must be momentum flowing into one part of the $GsG$ at the top.  If the diagram has no ``top'' interaction, such as the wrapped $uuuu$ structure, there must be one nonzero incoming momentum, in any propagator. If there is only one incoming momentum, we allow the operator to carry away the momentum $p$.  On the other hand, if, like in the unwrapped structure $udu$, there are two separate ``top'' interactions, the most convenient choice is to allow momentum $p$ flow into one, and momentum $-p$ flow into the other.  The diagrams $XuXdXuXd$ and $XuXuXuXu$ demonstrate this in Figure \ref{momenta}.

In either case the diagram can be expressed as a four-loop correction to a scalar propagator carrying some momentum $p$.  They take the form
$$
\lambda^{4}\int \frac{d^d q}{(2\pi)^d} \ \frac{d^d r}{(2\pi)^d} \ \frac{d^d s}{(2\pi)^d} \ \frac{d^d t}{(2\pi)^d} \Big( \cdots \Big) = \tilde{\lambda}^{4}\left(\frac{4\pi}{p^2}\right)^{4\epsilon}\Big(\cdots \Big)
$$
For the remainder of the calculations we will absorb the factor of $4\pi$ into  $p^2$ so that they need not be considered.

\subsection{Counterterm Subtraction Scheme}

As stated earlier, we renormalize $\mathcal{O}^{I} \rightarrow Z^{I}_{J}\mathcal{O}^{I}$, and $Z$ consists of an expansion in the coupling $\lambda$.  At some order $\lambda^{n}$ terms that contribute include products of $\lambda^{m}$ graphs in the bare correlator multiplied by $\lambda^{n-m}$ counterterms in the expansion of $Z$.  We can organize these terms conveniently by associating to each $\lambda^{n}$ graph those products of lower order graphs and counterterms which have the same structure; these we refer to as the ``daughter diagrams.''  While the divergent terms associated to a given diagram typically depend on the incoming momenta of the graph, we find that the divergent terms of the sum of a diagram and it's ``daughters'' is momentum independent-- it is a local counterterm.\footnote{Care should be taken that the daughter diagrams have the same incoming momenta as the original diagram.}

\begin{figure}
\begin{center}
\resizebox{5in}{!}{\includegraphics{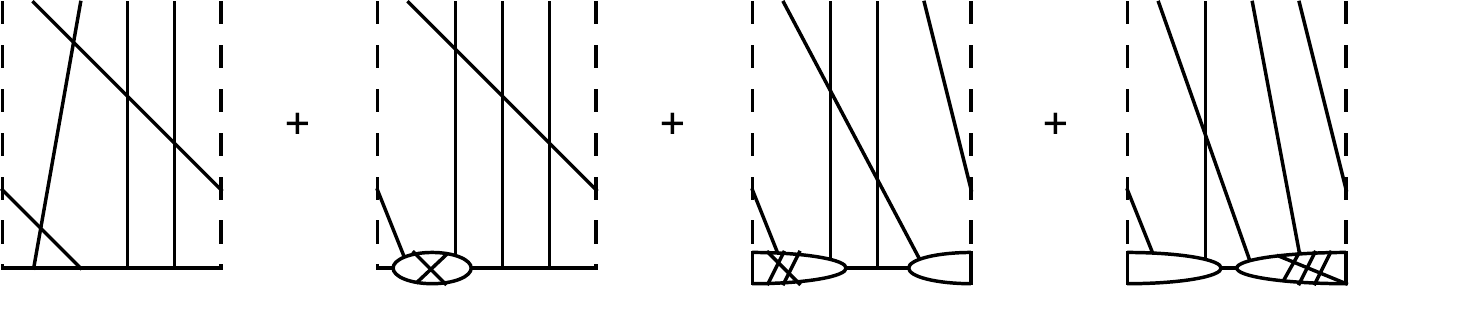}}
\end{center}
\caption{\label{XuXuXuXd} The sum of $XuXuXuXd$ and its daughter diagrams.}
\end{figure}

The daughter diagrams are obtained by consideration of the vertical ordering of interactions in a diagram.  In a diagram, any interaction or set of interactions which occur lower than all others may be ``dropped''  into a counterterm sitting at the bottom of the diagram.  The result is a daughter diagram where the dropped part is the counterterm and the remaining part is the lower-loop contribution to the correlator.  This is illustrated for diagram $XuXdXdXd$ in Figure \ref{XuXuXuXd}.  Using the prefix ``c'' to denote the counterterm of a diagram, we have
$$
cXuXuXuXd = - \mbox{div}\large(XuXdXdXd + cX\cdot XdXdX + cXdX\cdot XdX + cXdXdX \cdot X\large)
$$
where ``div'' is taken to mean that you extract only the divergent part of the expression.

\subsection{The Triangle Identity and Bubble Diagrams}

Having obtained diagrams of the form shown in Figure \ref{momenta}, we can use an integration by parts identity along with a general one-loop integral result to calculate most of the diagrams.  The ``triangle identity'' was worked out in \cite{triangle} and is also explained in \cite{BMR} and \cite{Smirnov:2004ym}.

\begin{figure}
\begin{center}
\resizebox{2in}{!}{\includegraphics{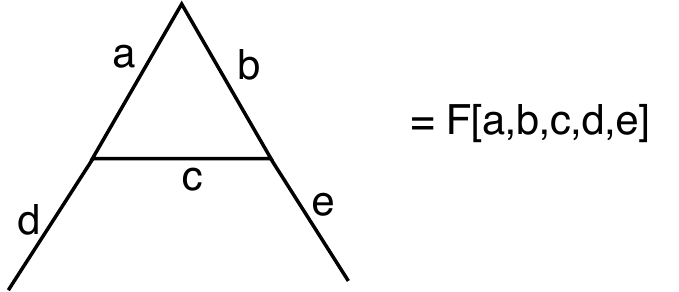}}
\end{center}
\caption{\label{triangle} The triangular loop that we can manipulate using partial integration}
\end{figure}

For a triangular loop of the form in Figure \ref{triangle} we have the relationship
\begin{eqnarray*}
(4 - 2\epsilon - a - b - 2c)F[a,b,c,d,e] & = & aF[a+1,b,c-1,d,e] - aF[a+1,b,c,d-1,e] \\
&& + bF[a,b+1,c-1,d,e] - bF[a,b+1,c,d,e-1]
\end{eqnarray*}
where the value of $a$ associated to a propagator indicates that the propagator $\frac{1}{q^2}$ should be raised to the power of $a$.

Most of the diagrams computed can be reduced by use of the triangle identity to products of ``bubble'' diagrams.  These are 1-loop corrections to a propagator where the two loop propagators are raised to arbitrary (not necessarily integer) powers.  These can be related to ratios of gamma functions by the usual Feynman parameter methods.

$$
L[a,b] = p^{2a + 2b - 4 + 2\epsilon}\int \frac{d^d q}{q^{2a}(q-p)^{2b}}
$$
$$
= \frac{\Gamma[a + b - 2 + \epsilon]\Gamma[2 - \epsilon + a]\Gamma[2 - \epsilon + b]}{\Gamma[a]\Gamma[b]\Gamma[4 - 2\epsilon - a - b]}
$$

\begin{figure}
\begin{center}
\resizebox{2in}{!}{\includegraphics{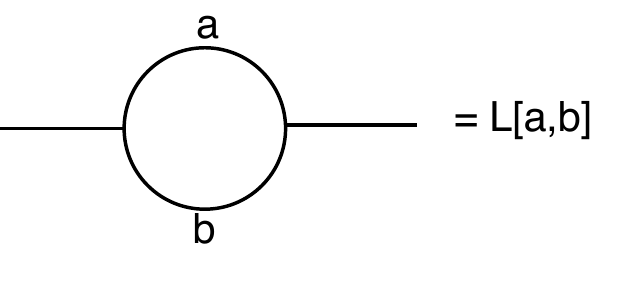}}
\end{center}
\caption{\label{L} The graphical representation for a bubble diagram.}
\end{figure}

\subsection{Irreducible Diagrams}

\begin{figure}
\begin{center}
\resizebox{6in}{!}{\includegraphics{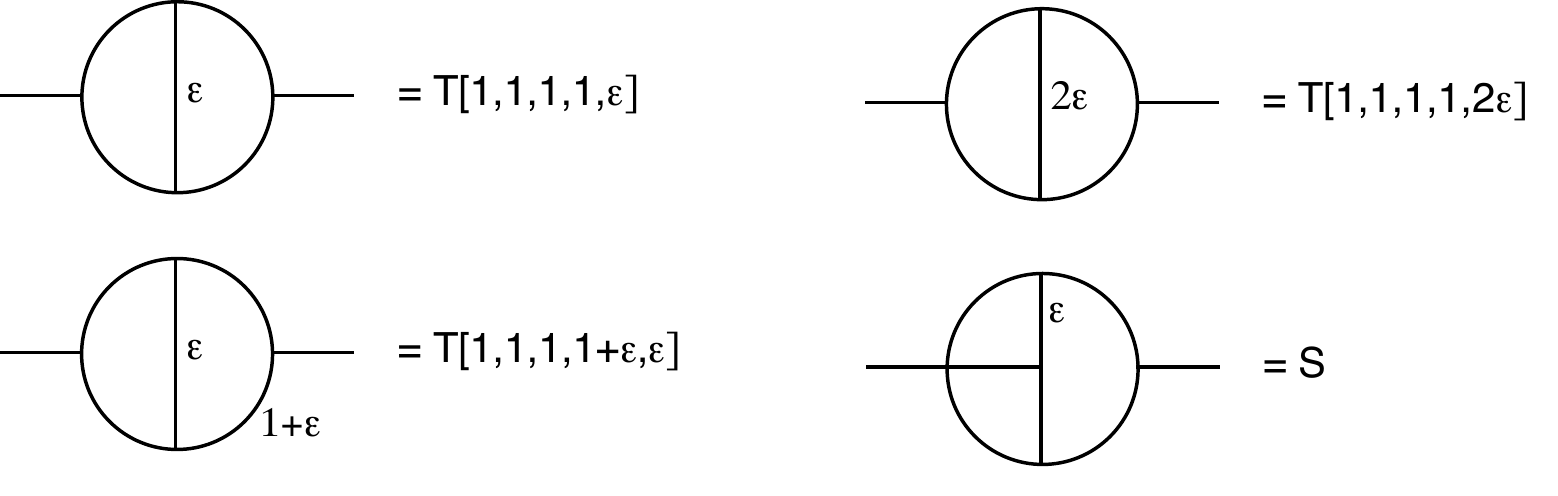}}
\end{center}
\caption{\label{ired} Diagrams that cannot be reduced to bubbles.}
\end{figure}

In some cases the diagrams calculated lead to subdiagrams that cannot be further simplified by use of the triangle identity.  There are four such subdiagrams that we encountered in the process of reducing the wrapped and unwrapped diagrams.  Three are are two-loop integrals of the type discussed in \cite{BMR}, and the fourth is a non-trivial three-loop integral.  These are shown in Figure \ref{ired}.

Although these diagrams cannot be computed in closed form, it is possible to generate a series expansion for them in $\epsilon$.    In our case, this was done by a process of ``reverse engineering''.  It is, as explained earlier, possible to compute the counterterm for a given diagram in different ways, by choosing different incoming momenta.  It is possible to find diagrams which can be reduced to bubbles using one choice of incoming momenta, but which involves a dependence on one of the above irriducible diagrams using a different choice.  Comparing the two results will give the first several terms in an expansion of the irreducible diagram.  (It is necessary when applying this technique to be sure that the daughter diagrams added to the main diagram have the appropriate incoming momenta for each reduction.)

\begin{figure}
\begin{center}
\resizebox{5in}{!}{\includegraphics{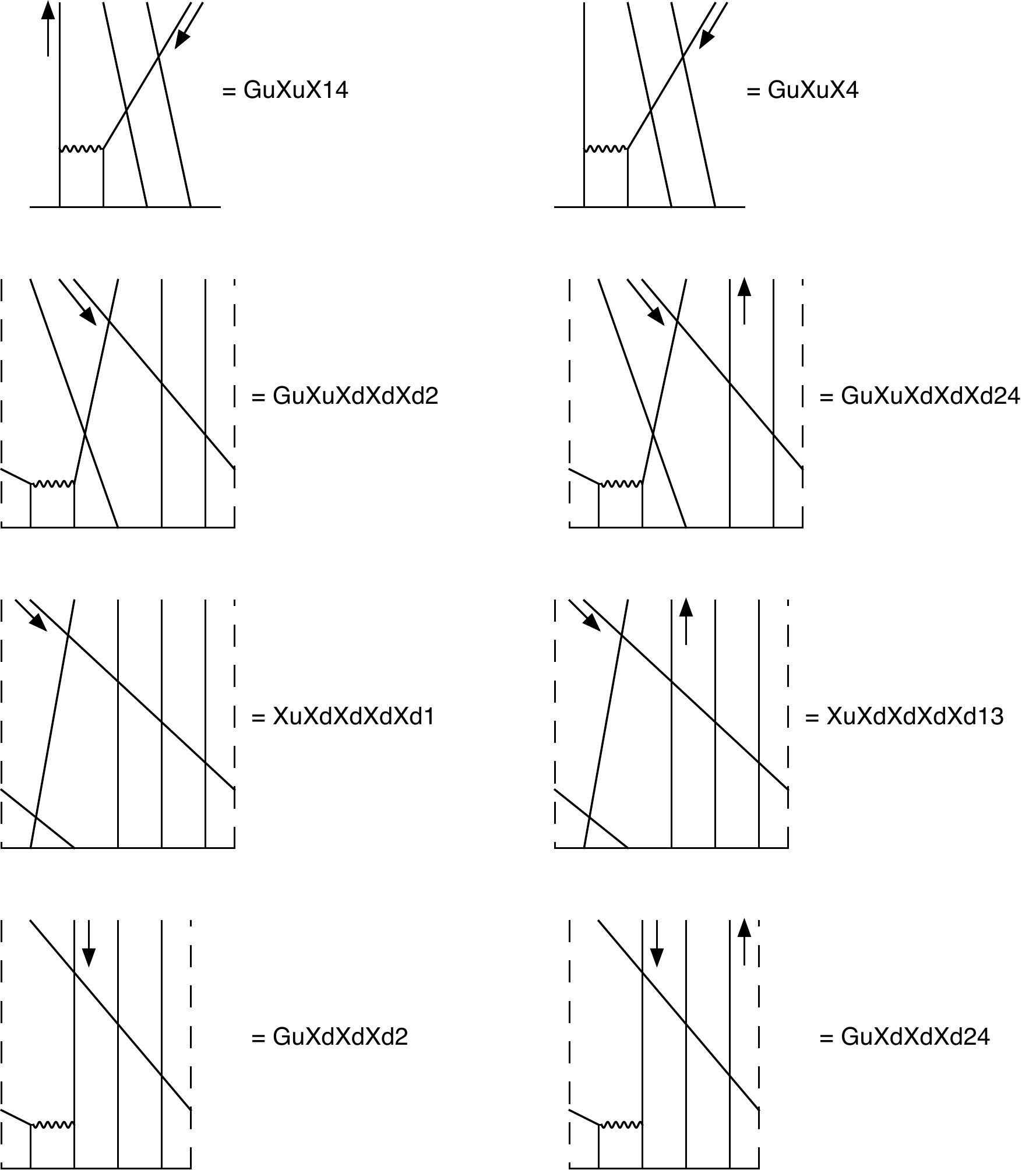}}
\end{center}
\caption{\label{iredmethod} Two different choices for incoming momenta used on diagrams $GuXuX$, $XuXdXdXdXd$, $GuXuXdXdXd$ and $GuXdXdXd$ labeled respectively by the legs on which momenta enter and exit.}
\end{figure}

In order to determine the first six terms in $T[1,1,1,1,\epsilon]$ (shown with the other irreducible diagrams in Figure \ref{ired}) we computed the counterterm associated with $GuXuX$ in two different ways, as shown in Figure \ref{iredmethod}.  For generating the first six terms in $T[1,1,1,1+\epsilon,\epsilon]$ we found it necessary to compute the five-loop wrapping diagram $XuXdXdXdXd$ in two ways.  For the first six terms in $T[1,1,1,1,2\epsilon]$ we computed the five-loop wrapping diagram $GuXdXdXdXd$, and for the first five terms in $S$ we compute $GuXdXdXd$.  Comparisons lead to the result

$$
T[1,1,1,1,\epsilon] = L[1,1]^2 \left\{\frac{1}{3} + \frac{\epsilon}{3} + \frac{\epsilon^2}{3} + \frac{7\epsilon^3(-1 + 2\zeta[3])}{3} + \epsilon^4\left(\frac{-67}{3} + \frac{7\pi^4}{90} + \frac{14\zeta[3]}{3}\right)  \right.
$$
$$
\left. +  \epsilon^5\left(\frac{-403}{3} + \frac{7\pi^4}{90} + \frac{86\zeta[3]}{3} + 126\zeta[5]\right)\right\}
$$

$$
T[1,1,1,1+\epsilon,\epsilon] = L[1,1]^2 \left\{\frac{5}{24} + \frac{5\epsilon}{12} + \frac{25\epsilon^2}{24} + \frac{\epsilon^3(5 + 38\zeta[3])}{12} + \epsilon^4\left(\frac{-455}{24} + \frac{19\pi^4}{360} + \frac{19\zeta[3]}{3}\right)  \right.
$$
$$
\left. +  \epsilon^5\left(\frac{-2215}{12} + \frac{19\pi^4}{180} + \frac{311\zeta[3]}{6} + \frac{341\zeta[5]}{2}\right)\right\}
$$

$$
T[1,1,1,1,2\epsilon] = L[1,1]^2\left\{\frac{1}{6} + \frac{\epsilon}{3} + \frac{\epsilon^2}{3} + \frac{\epsilon^3(-17 + 31\zeta[3])}{3} + \epsilon^4\left(\frac{-197}{3} + \frac{31\pi^4}{180} + \frac{62\zeta[3]}{3}\right) \right.
$$
$$
\left. + \epsilon^5\left(\frac{-1529}{3} + \frac{31\pi^4}{90} + \frac{386\zeta[3]}{3} + 449\zeta[5]\right)\right\}
$$
and
$$
S = L[1,1]^3\left\{\frac{1}{24} + \frac{\epsilon}{4} + \frac{37\epsilon^2}{24} + \frac{\epsilon^3(107 - 86\zeta[3])}{12} + \epsilon^4\left(\frac{1189}{24} - \frac{43\pi^4}{360} - 16\zeta[3]\right)\right\}.
$$

For further confirmation, these expansions were compared to numerical results obtained by using a Mellin-Barnes parametrization to write the integrals in terms of contour integrals in the complex plane \cite{Smirnov:2004ym}, where the algorithm for determining the coefficients of expansion in $\epsilon$ has been automated in \cite{MB}.\footnote{These numerical results were obtained with the help of Radu Roiban.}

\pagebreak

\subsection{Unwrapped Diagrams}

Here we present a complete list of the unwrapped, maximal length loop four counterterms.

$$
\begin{array}{ll}
cXuXuXuXu = \frac{1 + \epsilon(-6 + \epsilon(19 - 30\epsilon))}{24\epsilon^4} & \hspace{1in} cXuXuXdX = \frac{3 + \epsilon(-8 + \epsilon(5 + 8\epsilon))}{24\epsilon^4} \\
\\
cXuXdXuX = \frac{(-1+\epsilon)(-5 + \epsilon(5 + 6\epsilon))}{24\epsilon^4} & \hspace{1in} cXdXuXuX = \frac{3 + \epsilon(-12 + \epsilon(21 - 8\epsilon))}{24\epsilon^4} \\
 \\
 cXuXuXuG = \frac{1 + \epsilon(-6 + \epsilon(19 - 30\epsilon))}{24\epsilon^4} & \hspace{1in} cXuXuGuX = -\frac{(-1 + \epsilon)(1 + \epsilon(-3 + 8\epsilon))}{24\epsilon^4} \\
 \\
 cXuGuXuX = -\frac{(-1 + \epsilon)(1 + \epsilon(-1 + 6\epsilon))}{24\epsilon^4} & \hspace{1in} cGuXuXuX = \frac{1 + \epsilon^2(-5 + 8\epsilon)}{24\epsilon^4} \\
 \\
 cXuXuXdG = \frac{1 + \epsilon(-1 + 2\epsilon)(2 + 5\epsilon)}{8\epsilon^4} & \hspace{1in} cXuXuGdX = \frac{3 + \epsilon(-8 + \epsilon(5 + 8\epsilon))}{24\epsilon^4} \\
 \\
 cXuGuXdX = \frac{3 + \epsilon(-4 + \epsilon - 8\epsilon^2)}{24\epsilon^4} & \hspace{1in} cGuXuXdX = -\frac{(1 + \epsilon)(-3 + \epsilon + 6\epsilon^2)}{24\epsilon^4} \\
 \\
 cXuXdXuG = \frac{(-1 + \epsilon)(-5 + \epsilon(5 + 6\epsilon))}{24\epsilon^4} & \hspace{1in} cXuXdGuX = \frac{5 + \epsilon(4 + \epsilon(-17 + 8\epsilon))}{24\epsilon^4} \\
 \\
 cXuGdXuX = \frac{(-1 + \epsilon)(-5 + \epsilon(5 + 6\epsilon))}{24\epsilon^4} & \hspace{1in} cGuXdXuX = \frac{5 + \epsilon(-4 + \epsilon(-13 + 8\epsilon))}{24\epsilon^4} \\
 \\
 cXdXuXuG = \frac{3 + \epsilon(-12 + \epsilon(21 - 8\epsilon))}{24\epsilon^4} & \hspace{1in} cXdXuGuX = \frac{1 + \epsilon(-2 + \epsilon(3 - 2\epsilon))}{8\epsilon^4} \\
 \\
 cXdGuXuX = \frac{1 + \epsilon(2 + \epsilon(-5 + 2\epsilon))}{8\epsilon^4} & \hspace{1in} cGdXuXuX = \frac{3 + \epsilon(-12 + \epsilon(21 - 8\epsilon))}{24\epsilon^4} \\
 \\
 cXuXuGuG = -\frac{(-1 + \epsilon)(1 + \epsilon(-3 + 8\epsilon))}{24\epsilon^4} & \hspace{1in} cXuGuXuG =  -\frac{(-1 + \epsilon)(1 + \epsilon(-1 + 6\epsilon))}{24\epsilon^4}  \\
 \\
 cGuXuXuG = \frac{1 + \epsilon^2(-5 + 8\epsilon)}{24\epsilon^4} & \hspace{1in} cXuGuGuX = \frac{1 + \epsilon^2(7 + 8\epsilon)}{24\epsilon^4} \\
 \\
 cGuXuGuX = \frac{1 + \epsilon(2 + \epsilon(-1 + 6\epsilon))}{24\epsilon^4} & \hspace{1in} cGuGuXuX = \frac{1 + \epsilon(4 + \epsilon(7 - 8\epsilon))}{24\epsilon^4} \\
 \\
 cXuXuGsG = \frac{-1 + \epsilon(4 - 11\epsilon)}{24\epsilon^3} & \hspace{1in} cXuGsGuX = \frac{-1 + \epsilon - 4\epsilon^2}{12\epsilon^3} \\
 \\
 cGsGuXuX = \frac{-3 + \epsilon^2}{24\epsilon^3} & \hspace{1in} cXuXuGdG = \frac{1 + \epsilon(-1 + 2\epsilon)(2 + 5\epsilon)}{8\epsilon^4} \\
  \end{array}
 $$
 $$
\begin{array}{ll}
 cXuGuXdG = \frac{3 + \epsilon(-2 + \epsilon + 6\epsilon^2)}{24\epsilon^4} & \hspace{1in} cGuXuXdG = \frac{3 + \epsilon(4 - \epsilon(3 + 8\epsilon))}{24\epsilon^4} \\
 \\
 cXuGuGdX = \frac{3 + \epsilon(-4 + \epsilon - 8\epsilon^2)}{24\epsilon^4} & \hspace{1in} cGuXuGdX = -\frac{(1 + \epsilon)(-3 + \epsilon + 6\epsilon^2)}{24\epsilon^4} \\
 \\
 cGuGuXdX = \frac{1 - \epsilon(-2 + \epsilon + 10\epsilon^2)}{8\epsilon^4} & \hspace{1in} cXuGsGdX = \frac{-1 + \epsilon + 4\epsilon^2}{12\epsilon^3} \\
 \\
 cGsGuXdX = \frac{-5 + \epsilon(2 + 11\epsilon)}{24\epsilon^3} & \hspace{1in} cXuXdGuG = \frac{5 + \epsilon(-4 + \epsilon(-17 + 8\epsilon))}{24\epsilon^4} \\
 \\
 cXuGdXuG = \frac{(-1 + \epsilon)(-5 + \epsilon(5 + 6\epsilon))}{24\epsilon^4} &  \hspace{1in} cGuXdXuG = \frac{5 + \epsilon(-4 + \epsilon(-13 + 8\epsilon))}{24\epsilon^4} \\
\\
cXuGdGuX = \frac{5 + \epsilon(4 + \epsilon(-17 + 8\epsilon))}{24\epsilon^4} & \hspace{1in} cGuXdGuX = \frac{5 + \epsilon(10 + \epsilon(7 - 6\epsilon))}{24\epsilon^4} \\
\\
cGuGdXuX = \frac{5 + \epsilon(-4 + \epsilon(-13 + 8\epsilon))}{24\epsilon^4} & \hspace{1in} cXdXuGuG = \frac{1 + \epsilon(-2 + \epsilon(3 - 2\epsilon))}{8\epsilon^4} \\
 \\
 cXdGuXuG = \frac{1 + \epsilon(2 + \epsilon(-5 + 2\epsilon))}{8\epsilon^4} & \hspace{1in} cGdXuXuG = \frac{3 + \epsilon(-12 + \epsilon(21 - 8\epsilon))}{24\epsilon^4} \\
 \\
 cXdGuGuX = \frac{3 + \epsilon(12 + \epsilon(9 - 8\epsilon))}{24\epsilon^4} & \hspace{1in} cGdXuGuX = \frac{1 + \epsilon(-2 + \epsilon(3 - 2\epsilon))}{8\epsilon^4}  \\
 \\
 cGdGuXuX =  \frac{1 + \epsilon(2 + \epsilon(-5 + 2\epsilon))}{8\epsilon^4} & \hspace{1in} cXdXuGsG = -\frac{3 + \epsilon(-6 + \epsilon)}{24\epsilon^3} \\
 \\
 cXdGsGuX = \frac{-3 + \epsilon(-3 + 4\epsilon)}{12\epsilon^3} & \hspace{1in} cXuGuGuG = \frac{1 + \epsilon^2(7 + 8\epsilon)}{24\epsilon^4} \\
 \\
 cGuXuGuG = \frac{1 + \epsilon(2 + \epsilon(-1 + 6\epsilon))}{24\epsilon^4} & \hspace{1in} cGuGuXuG = \frac{1 + \epsilon(4 + \epsilon(7 - 8\epsilon))}{24\epsilon^4} \\
 \\
 cGuGuGuX = \frac{1 + \epsilon(6 + \epsilon(19 + 30\epsilon))}{24\epsilon^4} & \hspace{1in} cXuGuGsG = -\frac{1 + 7\epsilon^2}{24\epsilon^3} \\
 \\
 cXuGsGuG = \frac{-1 + \epsilon - 4\epsilon^2}{12\epsilon^3} & \hspace{1in} cGuXuGsG = \frac{-1 + \epsilon(-2 + \epsilon)}{24\epsilon^3} \\
 \\
 cGsGuXuG = \frac{-3 + \epsilon^2}{24\epsilon^3} & \hspace{1in} cGuGsGuX = -\frac{1 + \epsilon(5 + 6\epsilon)}{12\epsilon^3} \\
 \\
 cGsGuGuX = -\frac{3 + \epsilon(6 + 11\epsilon)}{24\epsilon^3} & \hspace{1in} cGsGsGuX = \frac{3 + \epsilon}{12\epsilon^2} \\
 \\
 cXuGsGsG = \frac{1 - \epsilon}{12\epsilon^2} & \hspace{1in} cXuGuGdG =  \frac{3 + \epsilon(-2 + \epsilon + 6\epsilon^2)}{24\epsilon^4} \\
 \\
 cGuXuGdG = \frac{3 + \epsilon(4 - \epsilon(3 + 8\epsilon))}{24\epsilon^4} & \hspace{1in} cGuGuXdG = \frac{3 + \epsilon(8 + \epsilon(9 + 8\epsilon))}{24\epsilon^4} \\
 \\
 cGuGuGdX = \frac{1 - \epsilon(-2 + \epsilon + 10\epsilon^2)}{8\epsilon^4} & \hspace{1in} cXuGsGdG = -\frac{1 + \epsilon - 6\epsilon^2}{12\epsilon^3} \\
 \\
 cGsGuXdG = -\frac{5 + \epsilon(4 + \epsilon)}{24\epsilon^3} & \hspace{1in} cGsGuGdX = \frac{-5 + \epsilon(2 + 11\epsilon)}{24\epsilon^3}  \\
  \end{array}
 $$
 $$
\begin{array}{ll}
 cXuGdGuG = \frac{5 + \epsilon(4 + \epsilon(-17 + 8\epsilon))}{24\epsilon^4} & \hspace{1in} cGuXdGuG = \frac{5 + \epsilon(10 + \epsilon(7 - 6\epsilon))}{24\epsilon^4} \\
 \\
 cGuGdXuG = \frac{5 + \epsilon(-4 + \epsilon(-13 + 8\epsilon))}{24\epsilon^4} & \hspace{1in} cGuGdGuX = \frac{5 + \epsilon(10 + \epsilon(7 - 6\epsilon))}{24\epsilon^4} \\
 \\
 cXdGuGuG = \frac{3 + \epsilon(12 + \epsilon(9 - 8\epsilon))}{24\epsilon^4} & \hspace{1in} cGdXuGuG = \frac{1 + \epsilon(-2 + \epsilon(3 - 2\epsilon))}{8\epsilon^4} \\
 \\
 cGdGuXuG = \frac{1 + \epsilon(2 + \epsilon(-5 + 2\epsilon))}{8\epsilon^4} & \hspace{1in} cGdGuGuX = \frac{3 + \epsilon(12 + \epsilon(9 - 8\epsilon))}{24\epsilon^4} \\
 \\
 cGsGdXuG = -\frac{3 + \epsilon(-6 + \epsilon)}{24\epsilon^3} & \hspace{1in} cXdGsGuG = \frac{-3 + \epsilon(-3 + 4\epsilon)}{12\epsilon^3} \\
 \\
 cGdXuGsG = -\frac{3 + \epsilon(-6 + \epsilon)}{24\epsilon^3} & \hspace{1in} cXdGuGsG = \frac{-3 + \epsilon(-12 + 7\epsilon)}{24\epsilon^3} \\
 \\
 cGsGdGuX = \frac{-3 + \epsilon(-12 + 7\epsilon)}{24\epsilon^3} & \\
\end{array}
$$

\subsection{Wrapped Diagrams}

Here we present a complete list of the four loop, length four, wrapped diagram counterterms.

$$
\begin{array}{ll}
cXuXuXuXu = \frac{-5\zeta[5]}{e} & \hspace{.75in} cXuXuXuXd = \frac{1 + \epsilon(-6 + \epsilon(19 + 6\epsilon(-5 + 4\zeta[3])))}{24\epsilon^4} \\
\\
cXuXuXdXd = \frac{1 + \epsilon(-4 + \epsilon(5 + 6\epsilon(1 - 2\zeta[3])))}{12\epsilon^4} & \hspace{.75in } cXuXdXuXd = \frac{1 - 2\epsilon(1 + \epsilon + 3\epsilon^2(-1 + \zeta[3]))}{6\epsilon^4} \\
\\
cXuXuXuGu = -\frac{\zeta[3]}{e} & \hspace{.75in} cXuXuXdGu = -\frac{-1 + \epsilon^2(5 + 4\epsilon(1 + 24\zeta[3] - 30\zeta[5]))}{24\epsilon^4} \\
\\
cXuXuXuGd = \frac{1 + \epsilon(-6 + \epsilon(19 + 6\epsilon(-5 + 4\zeta[3])))}{24\epsilon^4} & \hspace{.75in} cXuXuXdGd = \frac{1 + \epsilon(-3 + \epsilon + \epsilon^2(11 - 12\zeta[3]))}{12\epsilon^4} \\
\\
cXuXdXuGu = \frac{1 + \epsilon(-2 + \epsilon(7 + 6\epsilon(-3 + 4\zeta[3])))}{24\epsilon^4} & \hspace{.75in} cXuXdXdGu = -\frac{-1 + \epsilon(-2 + \epsilon(7 + 12\epsilon(1 + \zeta[3])))}{12\epsilon^4} \\
\\
cXuXdXuGd = \frac{1 - 2\epsilon(1 + \epsilon + 3\epsilon^2(-1 + \zeta[3]))}{6\epsilon^4} & \hspace{.75in} cXdXuXuGu = \frac{1 + \epsilon(-4 + \epsilon(11 + 4\epsilon(-5 + 6\zeta[3])))}{24\epsilon^4} \\
\\
cXdXuXdGu = \frac{1 - 2\epsilon^2(3 + \epsilon(1 + 3\zeta[3]))}{6\epsilon^4} & \hspace{.75in} cXdXuXuGd = \frac{1 + \epsilon(-4 + \epsilon(5 + 6\epsilon(1 - 2\zeta[3])))}{12\epsilon^4} \\
\\
cXuGuXuGu = \frac{1 - 2\zeta[3]}{2\epsilon} & \hspace{.75in} cXuGuXuGd = \frac{1 + \epsilon(-2 + \epsilon(7 + 6\epsilon(-3 + 4\zeta[3])))}{24\epsilon^4} \\
\\
cXuGuXdGu = \frac{1 + \epsilon(2 + \epsilon(-1 + 6\epsilon(5 + 4\zeta[3])))}{24\epsilon^4} & \hspace{.75in} cXuGuXdGd = \frac{1 + \epsilon(-2 + \epsilon + 6\epsilon^2(1 - 2\zeta[3]))}{12\epsilon^4} \\
\\
cXuGdXdGu = -\frac{-1 + \epsilon(-2 + \epsilon(7 + 12\epsilon(1 + \zeta[3])))}{12\epsilon^4} & \hspace{.75in} cXuGdXuGd = \frac{1 - 2\epsilon(1 + \epsilon + 3\epsilon^2(-1 + \zeta[3]))}{6\epsilon^4} \\
\\
\end{array}
$$
$$
\begin{array}{ll}
cXdGuXdGu = \frac{1 + 2\epsilon + 2\epsilon^2 + 6\epsilon^3(11 + 25\zeta[3] - 50\zeta[5])}{6\epsilon^4} & \hspace{.75in} cXuGuGuGu = \frac{3 - 2\zeta[3]}{2\epsilon} \\
\\
cXuGuGuGd = \frac{1 + \epsilon^2(7 + 4\epsilon(-7 + 6\zeta[3]))}{24\epsilon^4} & \hspace{.75in} cXuGuGdGu = \frac{1 + \epsilon(2 + \epsilon(-1 + 6\epsilon(5 +4\zeta[3])))}{24\epsilon^4} \\
\\
cXuGdGuGu = \frac{1 + \epsilon(4 + \epsilon(7 + 8\epsilon(2 + 3\zeta[3])))}{24\epsilon^4} & \hspace{.75in} cXdGuGuGu = \frac{1 + \epsilon(6 + \epsilon(19 - 6\epsilon(3 + 16\zeta[3] - 20\zeta[5])))}{24\epsilon^4} \\
\\
cXuGuGdGd = \frac{1 + \epsilon(-2 + \epsilon + 6\epsilon^2(1 - 2\zeta[3]))}{12\epsilon^4} & \hspace{.75in} cXuGdGdGu = \frac{1 + \epsilon(3 + \epsilon(-5 + \epsilon(5 + 48\zeta[3] - 60\zeta[5])))}{12\epsilon^4} \\
\\
cXdGdGuGu = \frac{1 + \epsilon(4 + \epsilon - 12\epsilon^2(1 + \zeta[3]))}{12\epsilon^4} & \hspace{.75in}  cXuGdGuGd = \frac{1 - 2\epsilon^2(3 + \epsilon(1 + 3\zeta[3]))}{6\epsilon^4} \\
\\
cXdGuGdGu = \frac{1 + 2\epsilon + 2\epsilon^2 + 6\epsilon^3(11 + 25\zeta[3] - 50\zeta[5])}{6\epsilon^4} & \hspace{.75in} cXuGuGsGu = \frac{-5}{4\epsilon} \\
\\
cXuGuGsGd = \frac{-1 + 5\epsilon^2}{24\epsilon^3} & \hspace{.75in} cXdGuGsGu = \frac{-1 - \epsilon(5 + 9\epsilon(7 + 14\zeta[3] - 30\zeta[5]))}{12\epsilon^3} \\
\\
cXuGdGsGd = \frac{-1 + \epsilon + 11\epsilon^2}{12\epsilon^3} & \hspace{.75in} cXdGdGsGu = -\frac{3 + \epsilon(6 + \epsilon(23 + 96\zeta[3] - 240\zeta[5]))}{24\epsilon^3} \\
\\
cXdGuGsGd = \frac{-1 - \epsilon(2 + \epsilon(17 + 60\zeta[3] - 60\zeta[5]))}{24\epsilon^3} & \hspace{.75in} cXdGdGsGd = \frac{-5}{4\epsilon} \\
\\
cXuGdGsGu = \frac{-3 + \epsilon^2(31 + 96\zeta[3] - 240\zeta[5])}{24\epsilon^3} & \hspace{.75in} cXuGsGsGu = \frac{-2\zeta[3] + 5\zeta[5]}{e} \\
\\
cXuGsGsGd = \frac{1 - 7\epsilon}{12\epsilon^2} & \hspace{.75in} cXdGsGsGu = \frac{3 - \epsilon(11 + 24\zeta[3] + 120\zeta[5])}{12\epsilon^2} \\
\\
cFcFcFbFb = \frac{2\zeta[3]}{e} & \hspace{.75in} cFcFcFcFb = \frac{\zeta[3] - 5\zeta[5]}{e} \\
\\
cFcFbFcFb = \frac{-2\zeta[3]}{e}
\end{array}
$$


\begin{thebibliography}{10}
\baselineskip=15pt

\bibitem{BES}
N.~Beisert, B.~Eden, and M.~Staudacher,
``Trancendentality and Crossing,''
J. Stat. Mech. {\bf 0701} (2007) P021
[arXiv:hep-th/0610251].

\bibitem{MZ}
J.~A.~Minahan and K.~Zarembo,
``The Bethe-Ansatz for $\mathcal{N} = 4$ Super Yang Mills,''
JHEP {\bf 0303} (2003) 013
[arXiv:hep-th/0212208].

\bibitem{background}
N.~Beisert, C.~Kristjansen and M.~Staudacher,
  ``The dilatation operator of N = 4 super Yang-Mills theory,''
  Nucl.\ Phys.\ B {\bf 664}, 131 (2003)
  [arXiv:hep-th/0303060].

N.~Beisert,
``The su(2$|$3) Dynamic Spin Chain,''
Nucl. Phys. {\bf B682} (2004) 487-520,
[arXiv:hep-th/0310252].

M.~Staudacher,
``The factorized S-matrix of CFT/AdS,''
JHEP {\bf 05} (2005) 054,
[arXiv:hep-th/0412188].

N.~Beisert, V.~Dippel, and M.~Staudacher,
``A novel long range spin chain and planar $\mathcal{N}=4$ super Yang-Mills'',
JHEP {\bf 07} (2004) 075,
[arXiv:hep-th/0405001].

R.~Hernandez and E.~Lopez,
``Quantum corrections to the string Bethe ansatz,''
JHEP {\bf 07} (2006) 004,
[arXiv:hep-th/0603204].

N.~Beisert, R.~Hernandez, and E.~Lopez,
``A crossing-symmetric phase for $\text{AdS}_5\times\text{S}^5$ strings,''
JHEP {\bf 11} (2006) 070,
[arXiv:hep-th/0609044].

\bibitem{Dixon}
Z.~Bern, M.~Czakon, L.~Dixon, D.~Kosower, and V.~Smirnov,
``The Four-Loop Planar Amplitude and Cusp Anomalous Dimension in Maximally Supersymmetric Yang-Mills Theory,''
Phys. Rev. {\bf D75} (2007) 085010,
[arXiv:hep-th/0610248].

F.~Cachazo, M.~Spradlin, and A.~Volovich,
``Four-Loop Cusp Anomalous Dimension from Obstructions,''
arXiv:hep-th0612309.

\bibitem{BMR}
N.~Beisert, T.~McLoughlin, and R.~Roiban,
``The Four-Loop Dressing Phase of $\mathcal{N}=4$ SYM,''
Phys. Rev. {\bf D76} (2007) 046002,
[arXiv:0705.0321[hep-th]].

\bibitem{wrapping}
C.~Sieg and A.~Torrielli,
``Wrapping interactions and the genus expansion of the 2- point function of composite operators,''
Nucl. Phys. {\bf B723} (2005) 3--32,
[arXiv:hep-th/0505071].

J.~Ambjorn, R.~Janik and C.~Kristjansen,
``Wrapping interactions and a new source of corrections to the spin-chain/string duality,''
Nucl. \ Phys. B \ {\bf 736} (2006) 288-301,
[hep-th/0510171].

\bibitem{talks}
N.~Mann
``Finite Size Corrections to the Bethe Ansatz in $\mathcal{N} = 4$ SYM:  A Diagrammatic Approach,''
talks given at Berkeley, October 2, Swansea, November 30, and Cambridge, December 10 (2007),
http://www.newton.cam.ac.uk/webseminars/pg+ws/2007/sis/sisw03/1210/mann/001.html

\bibitem{guesses}
A.~Rej, D.~Serban, and M.~Staudacher,
``Planar $\mathcal{N}=4$ gauge theory and the Hubbard model,''
JHEP {\bf 03} (2006) 018,
[arXiv:hep-th/0512077].

A.~V. Kotikov, L.~N. Lipatov, A.~Rej, M.~Staudacher, and V.~N. Velizhanin,
``Dressing and Wrapping,''
J. Stat. Mech. {\bf 0710} (2007) P10003,
[arXiv:0704.3586 [hep-th]].

\bibitem{Milano}
F.~Fiamberti, A.~Santambrogio, C.~Sieg, and D.~Zanon,
``Wrapping at Four Loops in $\mathcal{N} = 4$ SYM,''
arXiv:0712.3522 [hep-th].

\bibitem{ryzhov}
A.~V.~Ryzhov,
``Operators in the d = 4, N = 4 SYM and the AdS/CFT correspondence,''
arXiv:hep-th/0307169.

\bibitem{threeloops}
A.~V.~Kotikov, L.~N.~Lipatov, A.~I.~Onishchenko, and V.~N.~Velizhanin,
``Three-loop universal anomalous dimension of the Wilson operators in N = 4 SUSY Yang-Mills model,''
Phys. Lett. {\bf 595} (2004) 521-529,
[arXiv:hep-th/0404092].

B.~Eden, C.Jarczak, and E.~Sokatchev,
``A three-loop test of the dilatation operator in N = 4 SYM,''
Nucl. Phys. {\bf B712} (2005) 157-195,
[arXiv:hep-th/0409009].

\bibitem{Janik}
R.~A. Janik and T.~Lukowski,
''Wrapping interactions at strong coupling -- the giant magnon,''
Phys. Rev. {\bf D76} (2007) 126008,
[arXiv:0708.2208 [hep-th]].

\bibitem{KM2}
C.~Keeler and N.~Mann,
``Towards a finite size correction for General L,''
to be published.

\bibitem{GMR}
D.~Gross, A.~Mikhailov, and R.~Roiban,
``Operators with large R charge in N=4 Yang-Mills theory,''
Annals Phys. 301 (2002) 31-52,
[arXiv:hep-th/0205066].

\bibitem{EM}
  T.~G.~Erler and N.~Mann,
  ``Integrable open spin chains and the doubling trick in N = 2 SYM with
  fundamental matter,''
  JHEP {\bf 0601}, 131 (2006)
  [arXiv:hep-th/0508064].

\bibitem{triangle}
K.G.~Chetyrkin and F.V.~Tkachov,
``Integration by Parts: The Algorithm to Calculate beta Functions in 4 Loops,''
Nucl. Phys. {\bf B192} (1981) 159-204.

\bibitem{Smirnov:2004ym}
V.~A.~Smirnov,
``Evaluating Feynman Integrals,''
Springer Tracts Mod.\ Phys.\  {\bf 211}, 1 (2004).

\bibitem{MB}
C.~Anastasiou and A.~Daleo,
``Numerical evaluation of loop integrals,''
JHEP 0610, 031 (2006)
[arXiv:hep-ph/0511176].

\end{thebibliography}
\end{document}